# Mathematical Foundations of Quantum Computing for Computer Science Researchers

G. Fleury and P. Lacomme

*Université Clermont-Auvergne, Clermont-Auvergne-INP, LIMOS (UMR CNRS 6158),*
*1 rue de la Chebarde,*
*63178 Aubière Cedex, France*

[gerard.fleury@isima.fr](mailto:gerard.fleury@isima.fr), [philippe.lacomme@isima.fr](mailto:philippe.lacomme@isima.fr)

*Abstract*

This paper provides a short introduction to the mathematical foundation of quantum computation for researchers in computer science by providing an introduction fo the mathematical basis of calculations. This paper concerns the mathematical foundations of quantum computation addressing first the representation of qubit using the Bloch sphere and second the special relations between $SU(2)$ and $SO(3)$. The properties of $SU(2)$ are introduced focusing especially about the double-covering of $SO(3)$ and explaining how to map rotations of $SO(3)$ into matrices of $SU(2)$. Quantum physic operators are based on $SU(2)$ since we have a direct relationship to $SO(3)$ namely one isomorphism. We start first from basic representations of qubit in $\mathbb{R}^3$ and representations of operators in $SU(2)$ and we next discuss with operators that permit to move from one $SU(2)$ to another one according to a specific operator of $SU(2)$ that is related to rotation into $\mathbb{R}^3$.

**1. Introduction**

**1.1. $SU(2)$ definition**

Let us consider the $SU(n)$ group of $n \times n$ operators:

$$SU(n) = \{V \in GL(n, \mathbb{C}), V^\dagger = V^{-1} \text{ and } \det(V) = 1\}$$

To find the representation of $SU(2)$, let us consider $V \in SU(2)$ a general element of $SU(2)$ that can be written

$$V = \begin{pmatrix} a & b \\ c & d \end{pmatrix} \text{ and } V^\dagger = \begin{pmatrix} a^* & c^* \\ b^* & d^* \end{pmatrix} \text{ with } V^{-1} = \begin{pmatrix} d & -b \\ -c & a \end{pmatrix}$$

Since we have $V \in SU(2)$ it follows

$$V^\dagger = \begin{pmatrix} a^* & c^* \\ b^* & d^* \end{pmatrix} = \begin{pmatrix} d & -b \\ -c & a \end{pmatrix} = V^{-1}$$

This gives $a^* = d$, $b^* = -c$, $c^* = -b$ and $d^* = a$ and it follows that the representation is given by (considering that $c = -b^*$ and $d = a^*$):

$$V = \begin{pmatrix} a & b \\ c & d \end{pmatrix} = \begin{pmatrix} a & b \\ -b^* & a^* \end{pmatrix}$$

Moreover $\det V = 1$ hence $a.a^* + b.b^* = 1$ and if we assume $a = e + i.v$ and $b = h + i.m$ where $e, v, h, m \in \mathbb{R}$, then we have $e^2 + v^2 + h^2 + m^2 = 1$ that is the $S^3$ sphere of $\mathbb{R}^4$.

In the basis of Pauli Matrices, any operator $V \in SU(2)$ can be written $V = n_I.Id + n_X.i.X + n_Y.i.Y + n_Z.i.Z$

$$V = \begin{pmatrix} n_I + i.n_Z & i.n_X + n_y \\ i.n_X - n_y & n_I - i.n_Z \end{pmatrix} \text{ with } n_I^2 + n_X^2 + n_Y^2 + n_Z^2 = 1$$

Because $n_I^2 \leq 1$ we can assume $n_I = \cos\frac{\theta}{2}$. Then $n_X^2 + n_Y^2 + n_Z^2 = 1 - \cos^2\frac{\theta}{2}$ and we can assume $n_X^2 + n_Y^2 + n_Z^2 = \sin^2\frac{\theta}{2}$. So we can write:

$$V = \cos\frac{\theta}{2}.Id + i.\sin\frac{\theta}{2}.(x.X + y.Y + z.Z) \text{ with } x, y, z, \theta \in \mathbb{R}$$

Remark:

$$Z.Y = \begin{pmatrix} 1 & 0 \\ 0 & -1 \end{pmatrix}.\begin{pmatrix} 0 & -i \\ i & 0 \end{pmatrix} = \begin{pmatrix} 0 & -i \\ -i & 0 \end{pmatrix} = -i.X$$

$$Z.X = \begin{pmatrix} 1 & 0 \\ 0 & -1 \end{pmatrix}.\begin{pmatrix} 0 & 1 \\ 1 & 0 \end{pmatrix} = \begin{pmatrix} 0 & 1 \\ -1 & 0 \end{pmatrix} = i.Y$$

$$Y.Z = \begin{pmatrix} 0 & -i \\ i & 0 \end{pmatrix}.\begin{pmatrix} 1 & 0 \\ 0 & -1 \end{pmatrix} = \begin{pmatrix} 0 & i \\ i & 0 \end{pmatrix} = i.X$$

$$Y.X = \begin{pmatrix} 0 & -i \\ i & 0 \end{pmatrix}.\begin{pmatrix} 0 & 1 \\ 1 & 0 \end{pmatrix} = \begin{pmatrix} -i & 0 \\ 0 & i \end{pmatrix} = -i.Z$$

$$X.Z = \begin{pmatrix} 0 & 1 \\ 1 & 0 \end{pmatrix}.\begin{pmatrix} 1 & 0 \\ 0 & -1 \end{pmatrix} = \begin{pmatrix} 0 & -1 \\ 1 & 0 \end{pmatrix} = -i.Y$$

$$X.Y = \begin{pmatrix} 0 & 1 \\ 1 & 0 \end{pmatrix}.\begin{pmatrix} 0 & -i \\ i & 0 \end{pmatrix} = \begin{pmatrix} i & 0 \\ 0 & -i \end{pmatrix} = i.Z$$

$$\{X,Y\} = X.Y + Y.X = 0$$
$$\{X,Z\} = X.Z + Z.X = 0$$
$$\{Y,Z\} = Y.Z + Z.Y = 0$$

We must have

$$\cos^2\frac{\theta}{2}.Id + \sin^2\frac{\theta}{2}.(x^2 + y^2 + z^2 + x.y.XY + x.z.XZ + y.x.YX + y.z.Y.Z + z.x.Z.X + z.y.Z.Y) = 1$$

$$\cos^2\frac{\theta}{2}.Id + \sin^2\frac{\theta}{2}.(x^2 + y^2 + z^2) = 1$$

**Definition.**

$$\vec{n} \diamond \sigma = (n_x.X + n_y.Y + n_z.Z) = \begin{pmatrix} n_z & n_x - i.n_y \\ n_x + i.n_y & -n_z \end{pmatrix}$$

with $\sigma = (X, Y, Z)$ is the vector whose entries are the Pauli matrices while their dot product is not a scalar product but merely a convenient notation. We have $U(\vec{n}, \sigma) = i.\vec{n} \diamond \sigma$ the related $SU(2)$ operator to $\vec{n}$ with the explicit formulation:

$$U(\vec{n}, \sigma) = i.\vec{n} \diamond \sigma = i.\begin{pmatrix} n_z & n_x - i.n_y \\ n_x + i.n_y & -n_z \end{pmatrix}$$

☐

Any $V \in SU(2)$ can be written as :

$$V_{\vec{n},\theta} = \cos\left(\frac{\theta}{2}\right).\begin{pmatrix} 1 & 0 \\ 0 & 1 \end{pmatrix} + i.\sin\left(\frac{\theta}{2}\right).\begin{pmatrix} n_z & n_x - i.n_y \\ n_x + i.n_y & -n_z \end{pmatrix}$$

$$V_{\vec{n},\theta} = \begin{pmatrix} \cos\left(\frac{\theta}{2}\right) + i.n_z.\sin\left(\frac{\theta}{2}\right) & (i.n_x - n_y).\sin\left(\frac{\theta}{2}\right) \\ (i.n_x + n_y).\sin\left(\frac{\theta}{2}\right) & \cos\left(\frac{\theta}{2}\right) - i.n_z.\sin\left(\frac{\theta}{2}\right) \end{pmatrix} = \begin{pmatrix} a & b \\ -b^* & a^* \end{pmatrix}$$

with

$$a = \cos\left(\frac{\theta}{2}\right) + i.n_z.\sin\left(\frac{\theta}{2}\right)$$

$$a^* = \cos\left(\frac{\theta}{2}\right) - i.n_z.\sin\left(\frac{\theta}{2}\right)$$

$$b = (i.n_x - n_y).\sin\left(\frac{\theta}{2}\right)$$

$$-b^* = (i.n_x + n_y).\sin\left(\frac{\theta}{2}\right)$$

and $e = \cos\left(\frac{\theta}{2}\right)$, $v = n_z.\sin\left(\frac{\theta}{2}\right)$, $h = n_y.\sin\left(\frac{\theta}{2}\right)$ and $m = n_x.\sin\left(\frac{\theta}{2}\right)$.

Since $det(V) = 1$, we have:

$$e^2 + v^2 + h^2 + m^2 = 1$$

$$\cos\left(\frac{\theta}{2}\right)^2 + \sin\left(\frac{\theta}{2}\right)^2.n_x^2 + \sin\left(\frac{\theta}{2}\right)^2.n_y^2 + \sin\left(\frac{\theta}{2}\right)^2.n_z^2 = 1$$

$$\cos\left(\frac{\theta}{2}\right)^2 + \sin\left(\frac{\theta}{2}\right)^2.(n_x^2 + n_y^2 + n_z^2) = 1$$

So the result holds with $n_x^2 + n_y^2 + n_z^2 = 1$

**The following remarks hold:**

1) The function $\vec{n} \mapsto i.(\vec{n} \diamond \sigma)$ maps $\vec{n}$ and $i.\vec{n} \diamond \sigma$ and defines a bijection from $S^2$ and the subset of $SU(2)$ constituted by the family $i.(.\diamond \sigma)$.

2) The function $(\vec{n}, \theta) \mapsto V(\vec{n}, \theta)$ maps $S^2 \times [0; 4\pi[$ onto $SU(2)$ (Fig. 1).

$$V_{\vec{n},\theta} = \cos\left(\frac{\theta}{2}\right).\begin{pmatrix} 1 & 0 \\ 0 & 1 \end{pmatrix} + i.\sin\left(\frac{\theta}{2}\right).\begin{pmatrix} n_z & n_x - i.n_y \\ n_x + i.n_y & -n_z \end{pmatrix}$$

with $\theta \in [0; 4\pi[$

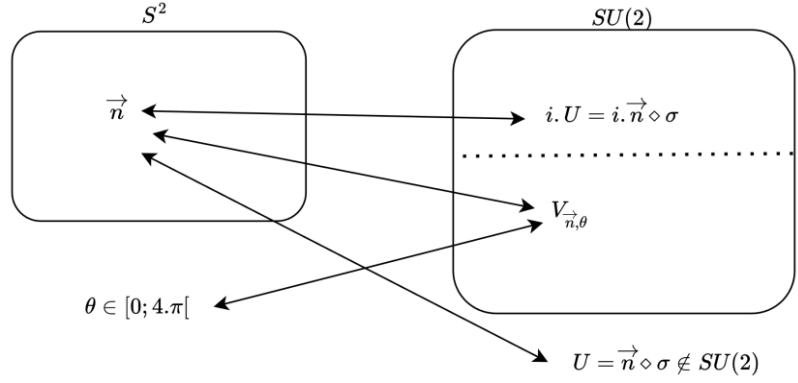

**Fig. 1.** Operators related to one $\vec{n}$ of $S^2$

Note that because $V_{-\vec{n},4\pi-\theta} = V_{-\vec{n},-\theta} = V_{\vec{n},\theta}$ we can consider only

$$V_{\vec{n},\theta} = \cos\left(\frac{\theta}{2}\right) \cdot \begin{pmatrix} 1 & 0 \\ 0 & 1 \end{pmatrix} + i \cdot \sin\left(\frac{\theta}{2}\right) \cdot \begin{pmatrix} n_z & n_x - i \cdot n_y \\ n_x + i \cdot n_y & -n_z \end{pmatrix}$$

with $\theta \in [0; 2\pi[$

Hence the function $(\vec{n}, \theta) \mapsto V_{\vec{n},\theta}$ from $S^2 \times [0; 2\pi[$ to $SU(2)$ is a bijection.

### 1.2. Mapping for $S^2$ and $SU(2)$

Any arbitrary vector $\vec{n}$ of $S^2$ such that $\vec{n} = (n_x, n_y, n_z)$ is related to the spherical coordinates based on 2 angles $\theta$ and $\phi$ (the spherical coordinates, also referred to as spherical polar coordinates, assume that $\theta \in [0; \pi]$ and $\phi \in [0; 2\pi[$) such the unit vector of $\mathbb{R}^3$ is parametrized through 2 real number $\theta$ and $\phi$ and illustrated in Fig. 2:

$$n_x = \sin\theta \cdot \cos\phi$$
$$n_y = \sin\theta \cdot \sin\phi \qquad (1)$$
$$n_z = \cos\theta$$

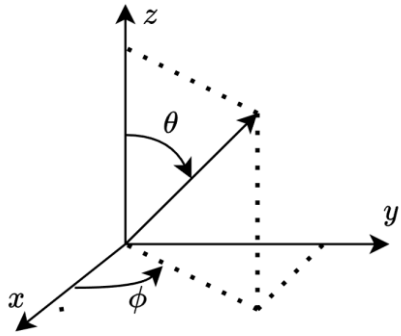

**Fig. 2.** Representation of one $\vec{n}$ of $S^2$

Inserting (1) into $U(\vec{n}, \sigma)$, the operator $U(\vec{n}, \sigma)$ can be rewritten:

$$U(\theta, \phi, \vec{n}) = i \cdot \vec{n} \diamond \sigma = i \cdot \begin{pmatrix} n_z & n_x - i \cdot n_y \\ n_x + i \cdot n_y & -n_z \end{pmatrix} = i \cdot \begin{pmatrix} \cos\theta & e^{-i\phi} \cdot \sin\theta \\ e^{i\phi} \cdot \sin\theta & -\cos\theta \end{pmatrix}$$

with $\theta \in [0; \pi]$ and $\phi \in [0; 2\pi[$

Because we have

$$U(\theta, \phi, -\vec{n}) = -i.\vec{n} \diamond \sigma = -i.\begin{pmatrix} n_z & n_x - i.n_y \\ n_x + i.n_y & -n_z \end{pmatrix} = i.\begin{pmatrix} -\cos\theta & -e^{-i.\phi}.\sin\theta \\ -e^{i.\phi}.\sin\theta & \cos\theta \end{pmatrix}$$

$$U(\pi - \theta, \pi + \phi, \vec{n}) = i.\begin{pmatrix} -\cos\theta & -e^{-i.\phi}.\sin\theta \\ -e^{i.\phi}.\sin\theta & \cos\theta \end{pmatrix} = U(\theta, \phi, -\vec{n})$$

the operator $U(\theta, \phi, \vec{n}) = i.\vec{n} \diamond \sigma$ can be defined considering only the angle $\theta \in [0; \pi]$ and $\phi \in [0; \pi[$.

**Notation**

$|\downarrow_u\rangle$ is the eigenvector corresponding to the eigenvalue $-1$

$|\uparrow_u\rangle$ is the eigenvector corresponding to the eigenvalue $+1$

☐

**Theorem**

The eigenvectors definition are $|\uparrow_{\vec{n}\diamond\sigma}\rangle = \begin{pmatrix} \cos\frac{\theta}{2} \\ e^{i.\phi}.\sin\frac{\theta}{2} \end{pmatrix}$ and $|\downarrow_{\vec{n}\diamond\sigma}\rangle = \begin{pmatrix} e^{-i.\phi}.\sin\frac{\theta}{2} \\ -\cos\frac{\theta}{2} \end{pmatrix}$ and $\vec{n} \diamond \sigma = |\uparrow_{\vec{n}\diamond\sigma}\rangle\langle\uparrow_{\vec{n}\diamond\sigma}| - |\downarrow_{\vec{n}\diamond\sigma}\rangle\langle\downarrow_{\vec{n}\diamond\sigma}|$. Moreover we have $\langle\uparrow_{\vec{n}\diamond\sigma}|\downarrow_{\vec{n}\diamond\sigma}\rangle = 0$ that is $|\uparrow_{\vec{n}\diamond\sigma}\rangle \perp |\downarrow_{\vec{n}\diamond\sigma}\rangle$ and we have $(\cos\theta \quad e^{-i.\phi}.\sin\theta) \cdot \begin{pmatrix} -e^{-i.\phi}.\sin\theta \\ \cos\theta \end{pmatrix} = -e^{-i.\phi}.\sin\theta.\cos\theta - e^{-i.\phi}.\sin\theta.\cos\theta = 0$.

**Proof.**

We have:

$$\vec{n} \diamond \sigma. |\uparrow_{\vec{n}\diamond\sigma}\rangle = \begin{pmatrix} \cos\theta & e^{-i.\phi}.\sin\theta \\ e^{+i.\phi}.\sin\theta & -\cos\theta \end{pmatrix} \begin{pmatrix} \cos\frac{\theta}{2} \\ e^{i.\phi}.\sin\frac{\theta}{2} \end{pmatrix} = \begin{pmatrix} \cos\theta.\cos\frac{\theta}{2} + e^{-i.\phi}.\sin\theta.e^{i.\phi}.\sin\frac{\theta}{2} \\ e^{+i.\phi}.\sin\theta.\cos\frac{\theta}{2} - e^{i.\phi}.\sin\frac{\theta}{2}\cos\theta \end{pmatrix}$$

$$\vec{n} \diamond \sigma. |\uparrow_{\vec{n}\diamond\sigma}\rangle = \begin{pmatrix} \cos\left(\theta - \frac{\theta}{2}\right) \\ e^{+i.\phi}\left[\sin\theta.\cos\frac{\theta}{2} - \sin\frac{\theta}{2}\cos\theta\right] \end{pmatrix} = \begin{pmatrix} \cos\left(\theta - \frac{\theta}{2}\right) \\ e^{+i.\phi}.\sin\left(\theta - \frac{\theta}{2}\right) \end{pmatrix}$$

And similarly

$$\vec{n} \diamond \sigma. |\downarrow_{\vec{n}\diamond\sigma}\rangle = \begin{pmatrix} \cos\theta & e^{-i.\phi}.\sin\theta \\ e^{+i.\phi}.\sin\theta & -\cos\theta \end{pmatrix} \begin{pmatrix} e^{-i.\phi}.\sin\frac{\theta}{2} \\ -\cos\frac{\theta}{2} \end{pmatrix} = \begin{pmatrix} \cos\theta.e^{-i.\phi}.\sin\frac{\theta}{2} - e^{-i.\phi}.\sin\theta.\cos\frac{\theta}{2} \\ e^{+i.\phi}.\sin\theta.e^{-i.\phi}.\sin\frac{\theta}{2} + \cos\frac{\theta}{2}\cos\theta \end{pmatrix}$$

$$\vec{n} \diamond \sigma. |\downarrow_{\vec{n}\diamond\sigma}\rangle = \begin{pmatrix} e^{-i.\phi}.\left[\cos\theta.\sin\frac{\theta}{2} - e^{-i.\phi}.\sin\theta.\cos\frac{\theta}{2}\right] \\ \sin\theta.\sin\frac{\theta}{2} + \cos\frac{\theta}{2}\cos\theta \end{pmatrix} = \begin{pmatrix} e^{-i.\phi}.\sin\left(-\frac{\theta}{2}\right) \\ \cos\frac{\theta}{2} \end{pmatrix} = -\begin{pmatrix} e^{-i.\phi}.\sin\frac{\theta}{2} \\ -\cos\frac{\theta}{2} \end{pmatrix}$$

When introducing $\vec{n} = (n_x, n_y, n_z) = (1,0,0)$ the operator $\vec{n} \diamond \vec{\sigma}$ becomes

$$\vec{n} \diamond \sigma = \begin{pmatrix} n_z & n_x - i.n_y \\ n_x + i.n_y & -n_z \end{pmatrix} = \begin{pmatrix} \cos\theta & e^{-i.\phi}.\sin\theta \\ e^{+i.\phi}.\sin\theta & -\cos\theta \end{pmatrix} = \begin{pmatrix} 0 & 1 \\ 1 & 0 \end{pmatrix}$$

with $n_x = \sin\theta.\cos\phi$, $n_y = \sin\theta.\sin\phi$ and $n_z = \cos\theta$

Because $n_z = \cos\theta = 1$ it gives $\theta = +\frac{\pi}{2}$ leading to

either $n_x = \cos\phi = 1$

$n_y = \sin\phi = 0$

$n_z = 0$ and $\theta = \frac{\pi}{2}$

Thus we can get away with only considering $\left(\theta = \frac{\pi}{2} \; ; \; \phi = 0\right)$ that coincides with the $x$-coordinates axis in the $\mathbb{R}^3$ in the case of Cartesian coordinates denoted $n_x, n_y$ and $n_z$.

The eigenvectors are defined by

$$|\uparrow_{\vec{n}\diamond\sigma}\rangle = |\uparrow_X\rangle = \begin{pmatrix} \cos\frac{\theta}{2} \\ e^{i.\phi}.\sin\frac{\theta}{2} \end{pmatrix} = \begin{pmatrix} \cos\frac{\pi}{4} \\ \sin\frac{\pi}{4} \end{pmatrix} = \begin{pmatrix} \frac{1}{\sqrt{2}} \\ \frac{1}{\sqrt{2}} \end{pmatrix} = \frac{1}{\sqrt{2}}.\begin{pmatrix} 1 \\ 1 \end{pmatrix}$$

$$|\downarrow_{\vec{n}\diamond\sigma}\rangle = |\downarrow_X\rangle = \begin{pmatrix} e^{-i.\phi}.\sin\frac{\theta}{2} \\ -\cos\frac{\theta}{2} \end{pmatrix} = \begin{pmatrix} \frac{1}{\sqrt{2}} \\ -\frac{1}{\sqrt{2}} \end{pmatrix} = \frac{1}{\sqrt{2}}.\begin{pmatrix} 1 \\ -1 \end{pmatrix}$$

which define the map between the $x$–axis of $S^2$ and the eigenvectors $|\uparrow_X\rangle$ and $|\downarrow_X\rangle$.

Introduction of $\vec{n} = (n_x, n_y, n_z) = (0,1,0)$ gives the operator $\vec{n} \diamond \sigma$ :

$$\vec{n} \diamond \sigma = \begin{pmatrix} n_z & n_x - i.n_y \\ n_x + i.n_y & -n_z \end{pmatrix} = \begin{pmatrix} \cos\theta & e^{-i.\phi}.\sin\theta \\ e^{+i.\phi}.\sin\theta & -\cos\theta \end{pmatrix} = \begin{pmatrix} 0 & -i \\ i & 0 \end{pmatrix}$$

with $n_x = \sin\theta.\cos\phi$, $n_y = \sin\theta.\sin\phi$ and $n_z = \cos\theta$

Because $n_z = \cos\theta = 0$ it gives $\theta = \frac{\pi}{2}$ and $n_y = \sin\phi = 1$ gives $\phi = \frac{\pi}{2}$.

Thus we can get away with only considering $\left(\theta = \frac{\pi}{2} \; ; \; \phi = \frac{\pi}{2}\right)$ that coincide with the y-coordinates axis in the $S^2$ in the case of Cartesian coordinates denoted $n_x, n_y$ and $n_z$ (Fig. 3).

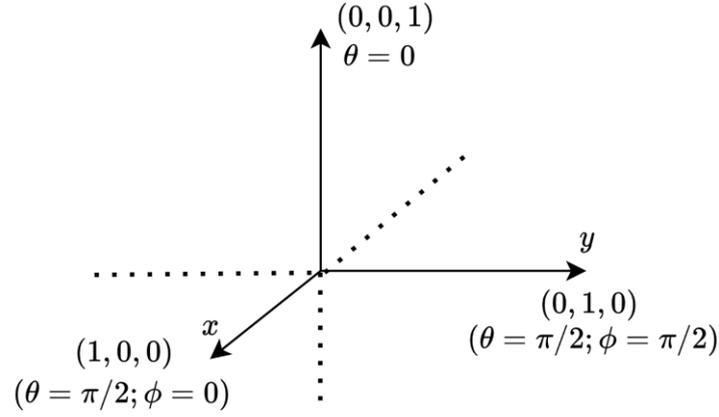

**Fig. 3.** Representation of one $\vec{n}$ of $S^2$

The eigenvectors are defined by

$$|\uparrow_u\rangle = \begin{pmatrix} \cos\frac{\theta}{2} \\ e^{i.\phi}.\sin\frac{\theta}{2} \end{pmatrix} = \begin{pmatrix} \cos\frac{\pi}{4} \\ e^{i.\frac{\pi}{2}}.\sin\frac{\pi}{4} \end{pmatrix} = \begin{pmatrix} \frac{1}{\sqrt{2}} \\ e^{i.\frac{\pi}{2}}.\sin\frac{\pi}{4} \end{pmatrix} = \begin{pmatrix} \frac{1}{\sqrt{2}} \\ i.\frac{1}{\sqrt{2}} \end{pmatrix}$$

$$|\downarrow_u\rangle = \begin{pmatrix} e^{-i.\phi}.\sin\frac{\theta}{2} \\ -\cos\frac{\theta}{2} \end{pmatrix} = \begin{pmatrix} e^{-i.\frac{\pi}{2}}.\sin\frac{\pi}{4} \\ -\cos\frac{\pi}{4} \end{pmatrix} = \begin{pmatrix} e^{-i.\frac{\pi}{2}}.\sin\frac{\pi}{4} \\ -\frac{1}{\sqrt{2}} \end{pmatrix} = \begin{pmatrix} i.\frac{1}{\sqrt{2}} \\ -\frac{1}{\sqrt{2}} \end{pmatrix}$$

defining the correspondence between the $y$ –axis of $S^2$ and the eigenvectors $|\uparrow_Y\rangle$ and $|\downarrow_Y\rangle$.

Introduction of $\vec{n} = (n_x, n_y, n_z) = (0,0,1)$ gives the operator $\vec{n} \diamond \sigma$ :

$$\vec{n} \diamond \sigma = \begin{pmatrix} n_z & n_x - i.n_y \\ n_x + i.n_y & -n_z \end{pmatrix} = \begin{pmatrix} \cos\theta & e^{-i.\phi}.\sin\theta \\ e^{+i.\phi}.\sin\theta & -\cos\theta \end{pmatrix} = \begin{pmatrix} 1 & 0 \\ 0 & -1 \end{pmatrix}$$

$$\vec{n} \diamond \sigma = \begin{pmatrix} \cos\theta & e^{-i.\phi}.\sin\theta \\ e^{+i.\phi}.\sin\theta & -\cos\theta \end{pmatrix}$$

with $n_x = \sin\theta.\cos\phi$, $n_y = \sin\theta.\sin\phi$ and $n_z = \cos\theta$

Because $n_z = \cos\theta = 1$ it gives $\theta = 0$ and $n_z = n_Y = 0$ with $\phi$ that is useless to check that ($\theta = 0$ ; $\phi$) coincide with the z-coordinates axis in the $S^2$ in the case of Cartesian coordinates denoted $n_x, n_y$ and $n_z$ (Fig. 2).

The eigenvectors are defined by

$$|\uparrow_{\vec{n}\diamond\sigma}\rangle = \begin{pmatrix} \cos\frac{\theta}{2} \\ e^{i.\phi}.\sin\frac{\theta}{2} \end{pmatrix} = \begin{pmatrix} \cos\frac{\pi}{4} \\ e^{i.\frac{\pi}{2}}.\sin\frac{\pi}{4} \end{pmatrix} = \begin{pmatrix} \frac{1}{\sqrt{2}} \\ e^{i.\frac{\pi}{2}}.\sin\frac{\pi}{4} \end{pmatrix} = \begin{pmatrix} 1 \\ 0 \end{pmatrix}$$

$$|\downarrow_{\vec{n}\diamond\sigma}\rangle = \begin{pmatrix} e^{-i.\phi}.\sin\frac{\theta}{2} \\ -\cos\frac{\theta}{2} \end{pmatrix} = \begin{pmatrix} e^{-i.\frac{\pi}{2}}.\sin\frac{\pi}{4} \\ -\cos\frac{\pi}{4} \end{pmatrix} = \begin{pmatrix} e^{-i.\frac{\pi}{2}}.\sin\frac{\pi}{4} \\ -\frac{1}{\sqrt{2}} \end{pmatrix} = \begin{pmatrix} 0 \\ 1 \end{pmatrix}$$

defining the correspondence between the $z$ –axis of $S^2$ and the eigenvectors $|\uparrow_Z\rangle$ and $|\downarrow_Z\rangle$.

We can state that identifying the $\vec{n}$ vector of $S^2$ to $\vec{n} \diamond \vec{\sigma}$ as gone from $S^2$ to $SU(2)$ and we have to show that the expected values of $|\uparrow_{\vec{n}\diamond\sigma}\rangle$ on the Pauli operators are mapped to $n_x, n_y, n_z$ and figure this out, we need to compute $\langle\uparrow_{\vec{n}\diamond\sigma}|\sigma|\uparrow_{\vec{n}\diamond\sigma}\rangle$.

For arbitrary $\vec{n}$ we get :

$$\langle\uparrow_{\vec{n}\diamond\sigma}|X|\uparrow_{\vec{n}\diamond\sigma}\rangle = \left(\cos\frac{\theta}{2}; e^{-i.\phi}.\sin\frac{\theta}{2}\right).\begin{pmatrix}0 & 1\\1 & 0\end{pmatrix}.\begin{pmatrix}\cos\frac{\theta}{2}\\e^{i.\phi}.\sin\frac{\theta}{2}\end{pmatrix}$$

$$\langle\uparrow_{\vec{n}\diamond\sigma}|X|\uparrow_{\vec{n}\diamond\sigma}\rangle = \cos\phi.\sin\theta = n_X$$

$$\langle\uparrow_{\vec{n}\diamond\sigma}|Y|\uparrow_{\vec{n}\diamond\sigma}\rangle = \left(\cos\frac{\theta}{2}; e^{-i.\phi}.\sin\frac{\theta}{2}\right).\begin{pmatrix}0 & -i\\i & 0\end{pmatrix}.\begin{pmatrix}\cos\frac{\theta}{2}\\e^{i.\phi}.\sin\frac{\theta}{2}\end{pmatrix}$$

$$\langle\uparrow_{\vec{n}\diamond\sigma}|Y|\uparrow_{\vec{n}\diamond\sigma}\rangle = \sin\phi.\sin\theta = n_Y$$

$$\langle\uparrow_{\vec{n}\diamond\sigma}|Z|\uparrow_{\vec{n}\diamond\sigma}\rangle = \left(\cos\frac{\theta}{2}; e^{-i.\phi}.\sin\frac{\theta}{2}\right).\begin{pmatrix}1 & 0\\0 & -1\end{pmatrix}.\begin{pmatrix}\cos\frac{\theta}{2}\\e^{i.\phi}.\sin\frac{\theta}{2}\end{pmatrix}$$

$$\langle\uparrow_{\vec{n}\diamond\sigma}|Z|\uparrow_{\vec{n}\diamond\sigma}\rangle = \cos\theta = n_Z$$

In a similar way

$$\langle\downarrow_{\vec{n}\diamond\sigma}|X|\downarrow_{\vec{n}\diamond\sigma}\rangle = \left(e^{i.\phi}.\sin\frac{\theta}{2}; -\cos\frac{\theta}{2}\right).\begin{pmatrix}0 & 1\\1 & 0\end{pmatrix}.\begin{pmatrix}e^{-i.\phi}.\sin\frac{\theta}{2}\\-\cos\frac{\theta}{2}\end{pmatrix}$$

$$\langle\downarrow_{\vec{n}\diamond\sigma}|X|\downarrow_{\vec{n}\diamond\sigma}\rangle = -e^{i.\phi}.\sin\frac{\theta}{2}.\cos\frac{\theta}{2} - e^{-i.\phi}.\sin\frac{\theta}{2}.\cos\frac{\theta}{2} = -\sin\frac{\theta}{2}.\cos\frac{\theta}{2}.[e^{i.\phi} + e^{-i.\phi}]$$

$$\langle\downarrow_{\vec{n}\diamond\sigma}|X|\downarrow_{\vec{n}\diamond\sigma}\rangle = -2.\sin\frac{\theta}{2}.\cos\frac{\theta}{2}.\cos\phi \text{ with } \sin\theta = 2.\sin\frac{\theta}{2}.\cos\frac{\theta}{2}$$

$$\langle\downarrow_{\vec{n}\diamond\sigma}|X|\downarrow_{\vec{n}\diamond\sigma}\rangle = -\sin\theta.\cos\phi = -n_X$$

And

$$\langle\downarrow_{\vec{n}\diamond\sigma}|Y|\downarrow_{\vec{n}\diamond\sigma}\rangle = \left(e^{i.\phi}.\sin\frac{\theta}{2}; -\cos\frac{\theta}{2}\right).\begin{pmatrix}0 & -i\\i & 0\end{pmatrix}.\begin{pmatrix}e^{-i.\phi}.\sin\frac{\theta}{2}\\-\cos\frac{\theta}{2}\end{pmatrix}$$

$$\langle \downarrow_{\vec{n}\diamond\sigma}|Y|\downarrow_{\vec{n}\diamond\sigma}\rangle = i.e^{i.\phi}.\sin\frac{\theta}{2}.\cos\frac{\theta}{2} - i.e^{-i.\phi}.\sin\frac{\theta}{2}.\cos\frac{\theta}{2}$$

$$\langle \downarrow_{\vec{n}\diamond\sigma}|Y|\downarrow_{\vec{n}\diamond\sigma}\rangle = i.\sin\frac{\theta}{2}.\cos\frac{\theta}{2}.[e^{i.\phi} - e^{-i.\phi}] = i.\sin\frac{\theta}{2}.\cos\frac{\theta}{2}.2.i.\sin\phi$$

$$\langle \downarrow_{\vec{n}\diamond\sigma}|Y|\downarrow_{\vec{n}\diamond\sigma}\rangle = -2.\sin\frac{\theta}{2}.\cos\frac{\theta}{2}.\sin\phi \text{ with } \sin\theta = 2.\sin\frac{\theta}{2}.\cos\frac{\theta}{2}$$

$$\langle \downarrow_{\vec{n}\diamond\sigma}|Y|\downarrow_{\vec{n}\diamond\sigma}\rangle = -\sin\theta.\sin\phi = -n_Y$$

And

$$\langle \downarrow_{\vec{n}\diamond\sigma}|Z|\downarrow_{\vec{n}\diamond\sigma}\rangle = \left(e^{i.\phi}.\sin\frac{\theta}{2}; -\cos\frac{\theta}{2}\right).\begin{pmatrix}1 & 0 \\ 0 & -1\end{pmatrix}.\begin{pmatrix}e^{-i.\phi}.\sin\frac{\theta}{2} \\ -\cos\frac{\theta}{2}\end{pmatrix}$$

$$\langle \downarrow_{\vec{n}\diamond\sigma}|Z|\downarrow_{\vec{n}\diamond\sigma}\rangle = \sin^2\frac{\theta}{2} - \cos^2\frac{\theta}{2} = -\cos\theta = -n_Z$$

Let a vector $\vec{n}$ of $S^2$ then a representation of the vector onto one operator with the standard Pauli matrices can the following map $\vec{n}\diamond\sigma$, and one $\vec{n}\diamond\sigma$ can be mapped to $S^2$ considering the expected values of the eigenvectors of $\vec{n}\diamond\vec{\sigma}$ on the Pauli matrices.

### 1.3. Geometric considerations

Now we can compute $|\langle\uparrow_k | \uparrow_{\vec{n}\diamond\sigma}\rangle|^2$ that is the probability outcomes of measurements made to $|k\rangle$.

Since we have $\langle\uparrow_X | \uparrow_{\vec{n}\diamond\sigma}\rangle = \left\langle\left(\frac{1}{\sqrt{2}};\frac{1}{\sqrt{2}}\right) \middle| \uparrow_{\vec{n}\diamond\sigma}\right\rangle = \left(\frac{1}{\sqrt{2}};\frac{1}{\sqrt{2}}\right).\begin{pmatrix}\cos\frac{\theta}{2} \\ e^{i.\phi}.\sin\frac{\theta}{2}\end{pmatrix} = \frac{1}{\sqrt{2}}.\cos\frac{\theta}{2} + \frac{1}{\sqrt{2}}.e^{i.\phi}.\sin\frac{\theta}{2}$ we get

$$|\langle\uparrow_X | \uparrow_{\vec{n}\diamond\sigma}\rangle|^2 = \left|\left(\frac{1}{\sqrt{2}}.\cos\frac{\theta}{2} + \frac{1}{\sqrt{2}}.e^{-i.\phi}.\sin\frac{\theta}{2}\right).\left(\frac{1}{\sqrt{2}}.\cos\frac{\theta}{2} + \frac{1}{\sqrt{2}}.e^{i.\phi}.\sin\frac{\theta}{2}\right)\right|^2 = \frac{1}{4}.(1 + \sin\theta.\cos\phi)^2$$

that models the probability outcomes of measurements made to $|\uparrow_X\rangle$.

This suggest that $|\langle\uparrow_X | \uparrow_{\vec{n}\diamond\sigma}\rangle|^2 = 1$ when $\sin\theta.\cos\phi = 1$ leading to $\sin\theta = \cos\phi = \pm 1$ i.e. $\sin\theta = \cos\phi = 1$ which proves that $\theta = \frac{\pi}{2}$ and $\phi = 0$ (since $\phi \in [0;\pi[$ the solution $\sin\theta = \cos\phi = -1$ i.e. $\theta = -\frac{\pi}{2}$ and $\phi = \pi$ is not relevant). In addition $|\langle\uparrow_X | \uparrow_{\vec{n}\diamond\sigma}\rangle|^2 = 0$ when $\sin\theta.\cos\phi = -1$ that leads to $\sin\theta = 1$ and $\cos\phi = -1$ i.e. $\theta = \frac{\pi}{2}$ and $\phi = \pi$ (the solution $\sin\theta = -1$ and $\cos\phi = 1$ that implies $\theta = \frac{3.\pi}{2}$ and $\phi = 0$ is not relevant since $\theta \in [0;\pi]$).

So the representation maps to the same vector space are equivalent and when the eigenvector $|\uparrow_{\vec{n}\diamond\sigma}\rangle$ is aligned to $|\uparrow_X\rangle$ it expresses the alignment of $\vec{n}$ with the $x-axis$ and when $|\uparrow_X\rangle$ and $|\uparrow_{\vec{n}\diamond\sigma}\rangle$ are orthogonal ($\langle\uparrow_X | \uparrow_{\vec{n}\diamond\sigma}\rangle = 0$) then we have align $\vec{n}$ in the opposite direction of the $x-axis$.

When $|\langle \uparrow_X | \uparrow_{\vec{n}\circ\sigma}\rangle|^2 = 0$, we have $\sin\theta . \cos\phi = -1$ leading to $\theta = \frac{\pi}{2}$ and $\phi \to \pi$ (since $\theta \in [0;\pi]$ the solution $\theta = -\frac{\pi}{2}$ and $\phi = 0$ is not relevant).

Similarly, we have

$$\langle \uparrow_y | \uparrow_u\rangle = \left\langle \left(\frac{i}{\sqrt{2}}; -\frac{1}{\sqrt{2}}\right) \middle| \uparrow_u\right\rangle = \left(\frac{i}{\sqrt{2}}; -\frac{1}{\sqrt{2}}\right) \cdot \begin{pmatrix} \cos\frac{\theta}{2} \\ e^{i.\phi}.\sin\frac{\theta}{2} \end{pmatrix} = \frac{i}{\sqrt{2}}.\cos\frac{\theta}{2} - \frac{1}{\sqrt{2}}.e^{i.\phi}.\sin\frac{\theta}{2}$$

and

$$|\langle \uparrow_y | \uparrow_u\rangle|^2 = \left(\frac{-i}{\sqrt{2}}.\cos\frac{\theta}{2} - \frac{1}{\sqrt{2}}.e^{-i.\phi}.\sin\frac{\theta}{2}\right) \cdot \left(\frac{i}{\sqrt{2}}.\cos\frac{\theta}{2} - \frac{1}{\sqrt{2}}.e^{i.\phi}.\sin\frac{\theta}{2}\right) = \frac{1}{2}.(1 + \sin\theta . \sin\phi)$$

and we have $|\langle \uparrow_Y | \uparrow_u\rangle|^2 = 1$ when $\sin\theta . \sin\phi = 1$ and $|\langle \uparrow_Y | \uparrow_u\rangle|^2 = 0$ when $\sin\theta . \sin\phi = -1$.

We have $|\langle \uparrow_Y | \uparrow_u\rangle|^2 = 1$ that lead to $\theta = \frac{\pi}{2}$ and $\phi = \frac{\pi}{2}$ ($\sin\theta = \sin\phi = -1$ i.e. $\theta = +\frac{3.\pi}{2}$ et $\phi = \frac{3.\pi}{2}$ is not relevant since $\theta \in [0; \pi[$) and we have $|\langle \uparrow_Y | \uparrow_u\rangle|^2 = 0$ that lead to $\theta = \frac{\pi}{2}$ et $\phi = -\frac{\pi}{2}$ ($\sin\theta = -1$ and $\sin\phi = 1$ and i.e. $\theta = +\frac{3.\pi}{2}$ et $\phi = \frac{\pi}{2}$ is not relevant since $\theta \in [0; \pi]$).

The eigenvector $|\uparrow_{\vec{n}\circ\sigma}\rangle$ is aligned to $|\uparrow_Y\rangle$ expresses the alignment of $\vec{n}$ with the $y-axis$ and when $|\uparrow_Y\rangle$ and $|\uparrow_{\vec{n}\circ\vec{\sigma}}\rangle$ are orthogonal ($\langle \uparrow_Y | \uparrow_{\vec{n}\circ\vec{\sigma}}\rangle = 0$) then we have align $\vec{n}$ in the opposite direction of the $y-axis$.

Let's consider $|\langle 0| \uparrow_{\vec{n}\circ\sigma}\rangle|^2 = \cos^2\frac{\theta}{2} = \frac{1}{2}.(1 + \cos\theta)$ that comes from the fact that $|\uparrow_Z\rangle = |0\rangle$ and $\langle 0| \uparrow_{\vec{n}\circ\sigma}\rangle = \cos\frac{\theta}{2}$ and which is the probability outcomes of measurements made to $|0\rangle$ and $\cos\theta$ is the dot product of $\vec{n}$ to $\vec{z}$ ($\cos\theta = \vec{n}.\vec{z}$).

For $|k\rangle = |\uparrow_Z\rangle$ we have $\langle \uparrow_Z | \uparrow_{\vec{n}\circ\sigma}\rangle = \langle (1;0)| \uparrow_{\vec{n}\circ\sigma}\rangle = (1;0) \cdot \begin{pmatrix} \cos\frac{\theta}{2} \\ e^{i.\phi}.\sin\frac{\theta}{2} \end{pmatrix} = \cos\frac{\theta}{2}$ and

$$|\langle \uparrow_Z | \uparrow_{\vec{n}\circ\sigma}\rangle|^2 = \cos^2\frac{\theta}{2} = \frac{1}{2}.(1 + \cos\theta)$$

that models the probability outcomes of measurements made to $|\uparrow_Z\rangle=|0\rangle$. As $\theta$ varies from 0 to $\pi$ this probability decreases from 1 to 0.

For any $\vec{k} \in S^2$ and $\vec{n} \in S^2$, we have $\vec{n}.\vec{k} = \langle \vec{n}|\vec{k}\rangle = \|\vec{n}\| + \|\vec{k}\| + \cos(\vec{n};\vec{k}) = \langle \vec{n}|\vec{k}\rangle = \cos(\vec{n};\vec{k})$ and $\langle \uparrow_k | \uparrow_n\rangle = \cos\frac{\vec{n}.\vec{k}}{2}$ considering $k = \vec{k} \diamond \sigma$ and $n = \vec{n} \diamond \sigma$ (Fig. 4)

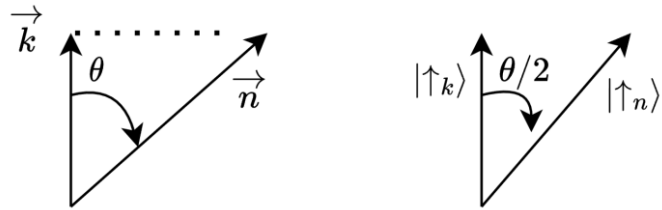

**Fig. 4.** Representation of one $\vec{n}$ of $S^2$

1.3. Principles

The above describe how the magnitude of an angle changes from 0 to $\pi$ and permits the definition of the three-dimensional Bloch (Nielsen and Chuang, 2010) sphere that provides a geometrical representation of a quantum state where the north and south are chosen to correspond to the standard basis vector $|\uparrow_Z\rangle = |0\rangle$ and $|\downarrow_Z\rangle = |1\rangle$. Note that $|\langle\uparrow_X | \uparrow_{\vec{n}\circ\sigma}\rangle|^2 = 0$ means $|\langle\downarrow_X | \uparrow_{\vec{n}\circ\sigma}\rangle|^2 = 1$, hence $|\uparrow_X\rangle \perp |\downarrow_X\rangle$ in $S^2$ but $|\uparrow_X\rangle = -|\downarrow_X\rangle$ in the Bloch sphere as shown by the values of $\phi, \theta$ angles. Similar remarks hold for $|\uparrow_Y\rangle$ and $|\downarrow_Y\rangle$ and for $|\uparrow_Z\rangle$ and $|\downarrow_Z\rangle$.

The eigenvectors of $-\vec{n} \circ \sigma$ and the eigenvectors $\vec{n} \circ \vec{\sigma}$ have been swapped i.e. $|\uparrow_{\vec{n}\circ\sigma}\rangle = |\downarrow_{-\vec{n}\circ\sigma}\rangle$. Note that we have $|\uparrow_{\vec{n}\circ\sigma}\rangle \perp |\downarrow_{\vec{n}\circ\sigma}\rangle$ and $|\uparrow_{\vec{n}\circ\sigma}\rangle \perp |\uparrow_{-\vec{n}\circ\sigma}\rangle$.

Such representation has been early used in numerous studies, providing a rigorous geometrical representation as stressed by (Feynmam et al., 1957) and Stroud speaks about Bloch vector in 1972 (Stroud, 1972). In the same year (Arecchi et al., 1972) introduce the Bloch state terminology (Fig. 5). The very first definition comes from Felix Bloch in his PhD dissertation (1929) who defines the mathematics of rotating coordinate procedures. One of the first representation of the Bloch sphere has been provided in 1974 by (Narducci et al., 1974). The eigenvectors $|\uparrow_X\rangle$ and $|\downarrow_X\rangle$ that are orthogonal and appear as opposite on the Bloch sphere and the same remark holds for $Y$ and $Z$.

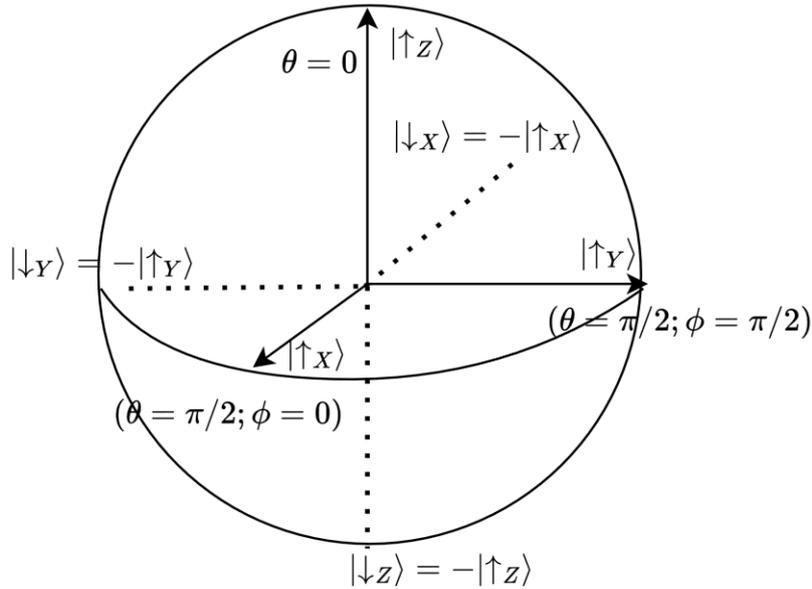

**Fig. 5.** The Bloch sphere

## 2. $SO(3)\ vs.\ SU(2)$

### 2.1. Principles

Once a Cartesian coordinate system has been chosen, the rotations in the three-dimensional space are represented by $3 \times 3$ real matrices $O$ such that $O.O^T = Id$ and such that $\det O = 1$.

**Notation**

$F_{\vec{n},\theta}(.)$ the rotation of $SO(3)$, through the angle $\theta$ about the direction $\vec{n}$ which is fully defined by the Rodrigues formula (Rodrigues, 1840) i.e.

$$\vec{x} \in S^2 \to R_{\vec{n},\theta}(\vec{x}) = \cos\theta \cdot \vec{x} + (1 - \cos\theta) \cdot \vec{x} \cdot \vec{n} + \sin\theta \cdot \vec{n} \wedge \vec{x}$$

A rotation $R_{\vec{n},\theta}(.)$ of $SO(3)$ transforms one unit vector $\vec{x} \in S^2$ into a new unit vector $\vec{x}' \in S^2$ as stressed in Fig 4. The previous section defined how to map any $\vec{n} \in S^2$ with $i.\vec{n} \diamond \sigma \in SU(2)$ considering:

$$i.\vec{n} \diamond \sigma = i.\begin{pmatrix} \cos\theta & e^{-i.\phi}.\sin\theta \\ e^{+i.\phi}.\sin\theta & -\cos\theta \end{pmatrix}$$

with $\vec{n}$ of $S^2$ such that $\vec{n} = (n_x, n_y, n_z)$ and

$$n_x = \sin\theta \cdot \cos\phi$$
$$n_y = \sin\theta \cdot \sin\phi$$
$$n_z = \cos\theta$$

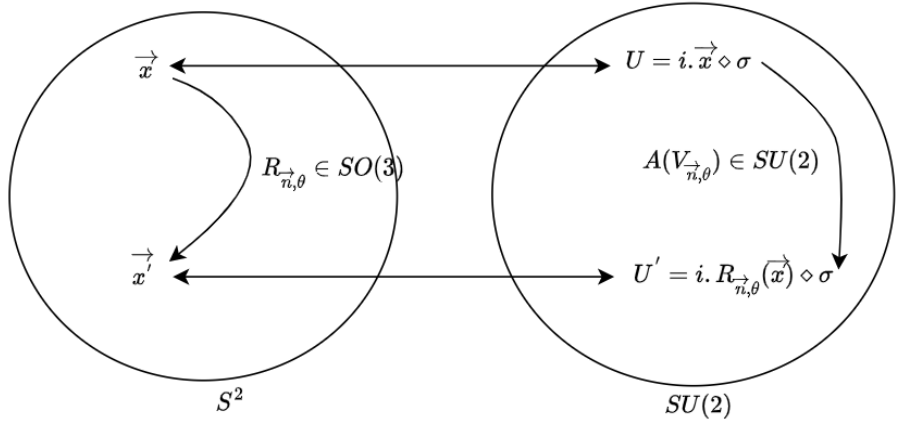

**Fig. 6.** Mapping between $S^2$ and $SU(2)$ and operators related to both spaces

**Theorem**

There exist

$$A(V_{\vec{n},\theta}) : SU(2) \to SU(2)$$
$$i.(. \diamond \sigma) \to i.R_{\vec{n},\theta}(.) \diamond \sigma$$

with

$$A(V_{\vec{n},\theta}) = V_{\vec{n},\theta} \cdot i.(. \diamond \sigma) \cdot (V_{\vec{n},\theta})^\dagger$$

and

$$V_{\vec{n},\theta} = \cos\left(\frac{\theta}{2}\right) \cdot Id + i.\sin\left(\frac{\theta}{2}\right) \cdot \vec{n} \diamond \vec{\sigma}$$

□

It defines an application $A(V_{\vec{n},\theta})$ (see Fig. 6) that is applied to one $i.(\vec{x} \diamond \sigma)$ (with $\vec{x} \in S^2$) to undergo a transformation of $i.(\vec{x} \diamond \sigma)$ into $i.(R_{\vec{n},\theta}(\vec{x}) \diamond \sigma)$ where $F_{\vec{n},\theta}$ is the rotation of $\vec{x}$ through the angle $\theta$ about the direction $\vec{n}$.

## 2.2. Properties of $R_{\vec{n},\theta}$

**Property** $R_{\vec{n},\theta}(k.\vec{n}) = k.\vec{n}$ with $k \in \mathbb{R}$

**Proof.**

If $(\vec{x} = k.\vec{n})$ $(\langle k.\vec{n}|\vec{n}\rangle = k)$ then

$$R_{\vec{n},\theta}(\vec{x}) = R_{\vec{n},\theta}(k.\vec{n}) = \cos\theta . k.\vec{n} + (1 - \cos\theta).k.\vec{n} + \sin\theta . \vec{n} \wedge (k.\vec{n})$$

$$R_{\vec{n},\theta}(\vec{x}) = k.\vec{n} + \sin\theta . \vec{0}$$

Because $\vec{n} \wedge (k.\vec{n}) = k.\vec{n} \wedge \vec{n} = \vec{0}$, we have

$$R_{\vec{n},\theta}(\vec{x}) = k.\vec{n} = \vec{x}$$

☐

**Theorem.** $R_{\vec{n},\theta}$ is the rotation through the angle $\theta$ about the direction $\vec{n}$ with $\theta \in ]-\pi; +\pi]$

**Proof.**

Considering any no-collinear vector $\vec{x}$ with $\vec{n}$ ($\vec{x}$ and $\vec{n}$ are linearly independent) and let us consider $\vec{v}$ that is spanned on $(\vec{x}, \vec{n})$ and that is orthogonal at $\vec{n}$, then $(\vec{n}, \vec{v}, \vec{n} \wedge \vec{v})$ is an orthonormal basis of $\mathbb{R}^3$.

Let us consider one element $\vec{x}$ of the plan $(\vec{x}, \vec{n})$ such that $\vec{x} = a.\vec{n} + b.\vec{v}$ (where $a$ and $b$ are real numbers).

We have:

$$R_{\vec{n},\theta}(\vec{x}) = \cos\theta . \vec{x} + (1 - \cos\theta).\langle\vec{x}|\vec{n}\rangle.\vec{n} + \sin\theta . \vec{n} \wedge \vec{x}$$

$$R_{\vec{n},\theta}(\vec{x}) = \cos\theta . \vec{x} + (1 - \cos\vartheta).\langle\vec{x}|\vec{n}\rangle.\vec{n} + \sin\theta . \vec{n} \wedge (a.\vec{n} + b.\vec{v})$$

$$R_{\vec{n},\theta}(\vec{x}) = \cos\theta . (a.\vec{n} + b.\vec{v}) + (1 - \cos\theta).a.\vec{n} + \sin\theta . b.\vec{n} \wedge \vec{v}$$

$$R_{\vec{n},\theta}(\vec{x}) = \cos\theta . b.\vec{v} + a.\vec{n} + \sin\theta . b.\vec{n} \wedge \vec{v}$$

And the norm is $\|R_{\vec{n},\theta}(\vec{x})\|^2 = |\cos\theta . b|^2 + |a|^2 + |\sin\theta . b|^2 = |a|^2 + |b|^2 = \|\vec{x}\|^2$ proving that $F$ is one isometry. A representation of $R_{\vec{n},\theta}(\vec{x})$ is given on Fig. 7.

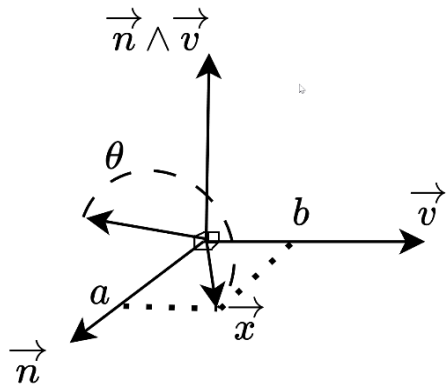

**Fig. 7.** Rotation through the angle $\theta$ about the direction $\vec{n}$

We note that applying the $R_{\vec{n},\theta}$ function to $\vec{x}$, the second component will correspond to $\cos\theta . \vec{v} + \sin\theta . \vec{n} \wedge \vec{v}$ that is a rotation i.e.:

$$R_{\vec{n},\theta}(\vec{x}) = a.\vec{n} + \cos\theta . b.\vec{v} + \sin\theta . b.\vec{n} \wedge \vec{v}$$

$$R_{\vec{n},\theta}(\vec{x}) = a.\vec{n} + b.(\cos\theta.\vec{v} + \sin\theta.\vec{n} \wedge \vec{v})$$

where $\cos\theta.\vec{v} + \sin\theta.\vec{n} \wedge \vec{v}$ is a rotation through the angle $\theta$ about the direction $\vec{n}$.

□

**Property.** $R_{-\vec{n},-\theta}(\vec{x}) = R_{\vec{n},\theta}$

**Proof.**

We have $R_{-\vec{n},-\theta}(\vec{x}) = -a.\vec{n} + b.(\cos\theta.\vec{v} + \sin\theta.\vec{n} \wedge \vec{v}) = R_{\vec{n},\theta}(\vec{x})$

So it is possible to consider only rotation through the angle $\theta$ about the direction $\vec{n}$ with $\theta \in [0;\pi]$

□

### 2.3. Preliminary results

Let one unit vector $\vec{n} \in S^2$ into with $\theta \in [0; 2\pi[$

Let us define

$$A(V_{\vec{n},\theta}) = V_{\vec{n},\theta}.i.(\diamond\,\sigma).(V_{\vec{n},\theta})^{\dagger}$$

With $V_{\vec{n},\theta} = \cos\left(\frac{\theta}{2}\right).Id + i.\sin\left(\frac{\theta}{2}\right).\vec{n} \diamond \vec{\sigma}$

$A(V_{\vec{n},\theta})$ is one operator of $SU(2)$ applied to $i.(\vec{x}.\diamond\,\sigma)$ another operator of $SU(2)$ that is mapped to $\vec{x}$ a vector of $S^2$:

$$A(V_{\vec{n},\theta})(\vec{x}) = \left(\cos\frac{\theta}{2}.Id + i.\sin\frac{\theta}{2}.\vec{n} \diamond \sigma\right).(i.(\vec{x} \diamond \sigma)).\left(\cos\frac{\theta}{2}.Id - i.\sin\frac{\theta}{2}.\vec{n} \diamond \sigma\right)$$

Since both $V_{\vec{n},\theta}$ and $i.(\diamond\,\sigma)$ are $SU(2)$ operators, then $A(V_{\vec{n},\theta})$ is a $SU(2)$ operator.

**Computation of $(\vec{x} \diamond \sigma).(\vec{n} \diamond \sigma)$**

$$(\vec{x} \diamond \sigma).(\vec{n} \diamond \sigma) = (x_1.X + x_2.Y + x_3.Z).(n_1.X + n_2.Y + n_3.Z)$$

$$(\vec{x} \diamond \sigma).(\vec{n} \diamond \sigma) = (x_1.n_1 + x_2.n_2 + x_3.n_3).Id + x_1.n_2.X.Y$$
$$+x_1.n_3.X.Z + x_2.n_1.Y.X + x_2.n_3.Y.Z + x_3.n_1.Z.X + x_3.n_2.Z.Y$$

$$(\vec{x} \diamond \sigma).(\vec{n} \diamond \sigma) = (x_1.n_1 + x_2.n_2 + x_3.n_3).Id + i.x_1.n_2.Z$$
$$-i.x_1.n_3.Y - i.x_2.n_1.Z + i.x_2.n_3.X + i.x_3.n_1.Y - i.x_3.n_2.X$$

$$(\vec{x} \diamond \sigma).(\vec{n} \diamond \sigma) = (x_1.n_1 + x_2.n_2 + x_3.n_3).Id + i.(x_2.n_3 - x_3.n_2).X$$
$$+i.(x_3.n_1 - x_1.n_3).Y + i.(x_1.n_2 - x_2.n_1).Z$$

$$(\vec{x} \diamond \sigma).(\vec{n} \diamond \sigma) = \langle\vec{x},\vec{n}\rangle.Id + i.(\vec{x} \wedge \vec{n}) \diamond \sigma$$

**Computation of $(\vec{n} \diamond \sigma).(\vec{x} \diamond \sigma)$**

$$(\vec{n} \diamond \sigma).(\vec{x} \diamond \sigma) = \langle\vec{x},\vec{n}\rangle.Id + i.(\vec{n} \wedge \vec{x}) \diamond \sigma = \langle\vec{x},\vec{n}\rangle.Id - i.(\vec{x} \wedge \vec{n}) \diamond \sigma$$

**Computation of $(\vec{n} \diamond \sigma).(\vec{n} \diamond \sigma)$**

$$(\vec{n} \diamond \sigma).(\vec{n} \diamond \sigma) = \langle\vec{n},\vec{n}\rangle.Id + i.(\vec{n} \wedge \vec{n}) \diamond \sigma = \langle\vec{n},\vec{n}\rangle.Id$$

**Computation of $(\vec{x} \diamond \sigma).(\vec{n} \diamond \sigma)$**

$$(\vec{x} \diamond \sigma).(\vec{n} \diamond \sigma) - (\vec{n} \diamond \sigma).(\vec{x} \diamond \sigma) = 2i.(\vec{x} \wedge \vec{n}) \diamond \sigma$$

**Computation of $(\vec{n} \diamond \sigma).(\vec{x} \diamond \sigma).(\vec{n} \diamond \sigma)$**

$$(\vec{n} \diamond \sigma).(\vec{x} \diamond \sigma).(\vec{n} \diamond \sigma) = (\vec{n} \diamond \sigma).[\langle \vec{x}, \vec{n} \rangle.Id + i.(\vec{x} \wedge \vec{n}) \diamond \sigma]$$

$$(\vec{n} \diamond \sigma).(\vec{x} \diamond \sigma).(\vec{n} \diamond \sigma) = \langle \vec{x}, \vec{n} \rangle.(\vec{n} \diamond \sigma).Id + i.(\vec{n} \diamond \sigma).[(\vec{x} \wedge \vec{n}) \diamond \sigma]$$

because we have $i.(\vec{n} \diamond \sigma).[(\vec{x} \wedge \vec{n}) \diamond \sigma] = -\bigl(\vec{n} \wedge (\vec{x} \wedge \vec{n})\bigr) \diamond \sigma$

we have:

$$(\vec{n} \diamond \sigma).(\vec{x} \diamond \sigma).(\vec{n} \diamond \sigma) = \langle \vec{x}, \vec{n} \rangle.(\vec{n} \diamond \sigma) - (\vec{n} \wedge (\vec{x} \wedge \vec{n})) \diamond \sigma$$

with $\vec{u} \wedge (\vec{v} \wedge \vec{w}) = \langle \vec{u}, \vec{w} \rangle.\vec{v} - \langle \vec{u}, \vec{v} \rangle.\vec{w}$

and $\vec{n} \wedge (\vec{x} \wedge \vec{n}) = \langle \vec{n}, \vec{n} \rangle.\vec{x} - \langle \vec{n}, \vec{x} \rangle.\vec{n}$

thus

$$(\vec{n} \diamond \sigma).(\vec{x} \diamond \sigma).(\vec{n} \diamond \sigma) = \langle \vec{x}, \vec{n} \rangle.\vec{n} \diamond \sigma.Id - [\langle \vec{n}, \vec{n} \rangle.\vec{x} - \langle \vec{n}, \vec{x} \rangle.\vec{n}] \diamond \sigma$$

$$(\vec{n} \diamond \sigma).(\vec{x} \diamond \sigma).(\vec{n} \diamond \sigma) = \langle \vec{x}, \vec{n} \rangle.\vec{n} \diamond \sigma - \langle \vec{n}, \vec{n} \rangle.\vec{x} \diamond \sigma + \langle \vec{n}, \vec{x} \rangle.\vec{n} \diamond \sigma$$

$$(\vec{n} \diamond \sigma).(\vec{x} \diamond \sigma).(\vec{n} \diamond \sigma) = 2.\langle \vec{x}, \vec{n} \rangle.\vec{n} \diamond \sigma - \langle \vec{n}, \vec{n} \rangle.\vec{x} \diamond \sigma$$

## 2.4. Definition of $A(V_{\theta,\vec{n}})$

It defines an application of $A(V_{\theta,\vec{n}})$ that is applied to one vector $\vec{x} \in S^2$ undergoes a rotation of $\vec{x}$ through the angle $\theta$ about the direction $\vec{n}$.

$$A(V_{\vec{n},\theta})(\vec{x}) = i.\left(\cos\frac{\theta}{2}.Id + i.\sin\frac{\theta}{2}.\vec{n} \diamond \sigma\right).(\vec{x} \diamond \sigma).\left(\cos\frac{\theta}{2}.Id - i.\sin\frac{\theta}{2}.\vec{n} \diamond \sigma\right)$$

$$A(V_{\vec{n},\theta})(\vec{x}) = i.\left(\cos\frac{\theta}{2}.Id + i.\sin\frac{\theta}{2}.\vec{n} \diamond \sigma\right).\cos\frac{\theta}{2}.(\vec{x} \diamond \sigma)$$
$$- i.\left(\cos\frac{\theta}{2}.Id + i.\sin\frac{\theta}{2}.\vec{n} \diamond \sigma\right).i.\sin\frac{\theta}{2}.(\vec{x} \diamond \sigma).(\vec{n} \diamond \sigma)$$

$$A(V_{\vec{n},\theta})(\vec{x}) = i.\left[\cos^2\frac{\theta}{2}.(\vec{x} \diamond \sigma) + i.\cos\frac{\theta}{2}.\sin\frac{\theta}{2}.(\vec{n} \diamond \sigma).(\vec{x} \diamond \sigma)\right.$$
$$\left. - i.\sin\frac{\theta}{2}.\left(\cos\frac{\theta}{2}.Id + i.\sin\frac{\theta}{2}.\vec{n} \diamond \sigma\right).(\vec{x} \diamond \sigma).(\vec{n} \diamond \sigma)\right]$$

$$A(V_{\vec{n},\theta})(\vec{x}) = i.\left[\cos^2\frac{\theta}{2}.(\vec{x} \diamond \sigma) + i.\cos\frac{\theta}{2}.\sin\frac{\theta}{2}.(\vec{n} \diamond \sigma).(\vec{x} \diamond \sigma) - i.\sin\frac{\theta}{2}.\cos\frac{\theta}{2}.(\vec{x} \diamond \sigma).(\vec{n} \diamond \sigma)\right.$$
$$\left. + \sin^2\frac{\theta}{2}.(\vec{n} \diamond \sigma).(\vec{x} \diamond \sigma).(\vec{n} \diamond \sigma)\right]$$

$$A(V_{\vec{n},\theta})(\vec{x}) = i.\left[\cos^2\frac{\theta}{2}.\vec{x} \diamond \sigma - i.\cos\frac{\theta}{2}.\sin\frac{\theta}{2}.[(\vec{x} \diamond \sigma).(\vec{n} \diamond \sigma) - (\vec{n} \diamond \sigma).(\vec{x} \diamond \sigma)]\right.$$
$$\left. + \sin^2\frac{\theta}{2}.(\vec{n} \diamond \sigma).(\vec{x} \diamond \sigma).(\vec{n} \diamond \sigma)\right]$$

with $(\vec{x} \diamond \sigma).(\vec{n} \diamond \sigma) - (\vec{n} \diamond \sigma).(\vec{x} \diamond \sigma) = 2i.(\vec{x} \wedge \vec{n}) \diamond \sigma$

and $(\vec{n} \diamond \sigma).(\vec{x} \diamond \sigma).(\vec{n} \diamond \sigma) = 2.\langle \vec{x}, \vec{n} \rangle.\vec{n} \diamond \sigma - \langle \vec{n}, \vec{n} \rangle.\vec{x} \diamond \sigma$

thus

$$A(V_{\vec{n},\theta})(\vec{x}) = i.\left[\cos^2\frac{\theta}{2}.\vec{x} \diamond \sigma - i.\cos\frac{\theta}{2}.\sin\frac{\theta}{2}.2i.(\vec{x} \wedge \vec{n}) \diamond \sigma + \sin^2\frac{\theta}{2}.[2.\langle\vec{x},\vec{n}\rangle.\vec{n} \diamond \sigma - \langle\vec{n},\vec{n}\rangle.\vec{x} \diamond \sigma]\right]$$

$$A(V_{\vec{n},\theta})(\vec{x}) = i.\left[\cos^2\frac{\theta}{2}.(\vec{x} \diamond \sigma) + 2.\cos\frac{\theta}{2}.\sin\frac{\theta}{2}.(\vec{x} \wedge \vec{n}) \diamond \sigma + 2.\sin^2\frac{\theta}{2}.\langle\vec{x},\vec{n}\rangle.(\vec{n} \diamond \sigma)\right.$$
$$\left. - \sin^2\frac{\theta}{2}.\langle\vec{n},\vec{n}\rangle.(\vec{x} \diamond \sigma)\right]$$

$$A(V_{\vec{n},\theta})(\vec{x}) = i.\left[\left(\cos^2\frac{\theta}{2} - \sin^2\frac{\theta}{2}.\langle\vec{n},\vec{n}\rangle\right).(\vec{x} \diamond \sigma) + 2.\cos\frac{\theta}{2}.\sin\frac{\theta}{2}.(\vec{x} \wedge \vec{n}) \diamond \sigma + 2.\sin^2\frac{\theta}{2}.\langle\vec{x},\vec{n}\rangle.(\vec{n} \diamond \sigma)\right]$$

$$A(V_{\vec{n},\theta})(\vec{x}) = i.[\cos\theta.(\vec{x} \diamond \sigma) + \sin\theta.(\vec{x} \wedge \vec{n}) \diamond \sigma + (1 - \cos\theta).\langle\vec{x},\vec{n}\rangle.(\vec{n} \diamond \sigma)]$$

$$A(V_{\vec{n},\theta})(\vec{x}) = i.[\cos\theta.\vec{x} + (1 - \cos\theta).\langle\vec{x},\vec{n}\rangle.\vec{n} + \sin\theta.(\vec{x} \wedge \vec{n})] \diamond \sigma$$

$$A(V_{\vec{n},\theta})(\vec{x}) = i.R_{\vec{n},\theta}(\vec{x}) \diamond \sigma$$

□

If $\theta = 0$ then $V_{\vec{n},0} = Id$ and we have

$$A_{\vec{n},0}(\vec{x} \diamond \sigma) = [\cos 0.\vec{x} + (1 - \cos 0).\langle\vec{x},\vec{n}\rangle.\vec{n} - \sin 0.(\vec{x} \wedge \vec{n})] \diamond \sigma$$

$$A_{\vec{n},0}(\vec{x} \diamond \sigma) = \vec{x} \diamond \sigma \text{ i.e. } A_{\vec{n},0} = Id$$

If $\theta = 2.\pi$ then $V_{\vec{n},2\pi} = (\cos\pi.Id - i.\sin\pi.\vec{n} \diamond \sigma) = -Id$ and we have

$$A_{\vec{n},2\pi}(\vec{x} \diamond \sigma) = [\cos 2\pi.\vec{x} + (1 - \cos 2\pi).\langle\vec{x},\vec{n}\rangle.\vec{n} - \sin 2\pi.(\vec{x} \wedge \vec{n})] \diamond \sigma$$

$$A_{\vec{n},2\pi}(\vec{x} \diamond \sigma) = \vec{x} \diamond \sigma \text{ i.e. } A_{\vec{n},2\pi} = Id$$

**Conclusion**

Because $A_{\vec{n},2\pi}(\vec{x} \diamond \sigma) = Id$, it is possible to define

$$A_{\vec{n},\theta}(\vec{x} \diamond \sigma) = \cos\theta.(\vec{x} \diamond \sigma) + (1 - \cos\theta).\langle\vec{x},\vec{n}\rangle.(\vec{n} \diamond \sigma) - \sin\theta.(\vec{x} \wedge \vec{n}) \diamond \sigma \text{ with } \theta \in [0; 2\pi[$$

Meaning that any operator $V_{\vec{n},\theta} \in SU(2)$ defined by $V_{\vec{n},\theta} = \cos\left(\frac{\theta}{2}\right).Id. + \sin\left(\frac{\theta}{2}\right).i.\vec{n} \diamond \vec{\sigma}$ (with $\|\vec{n}\| = 1$ and $\theta \in [0; 2\pi[$) defines a rotation $R_{\vec{n},\theta} \in SO(3)$ (with $\|\vec{n}\| = 1$ et $\theta \in [0; 2\pi[$).

We have:

$$A_{\vec{n},2\pi-\theta}(\vec{x} \diamond \sigma) = \cos(2\pi - \theta).\vec{x} + (1 - \cos(2\pi - \theta)).\langle\vec{x},\vec{n}\rangle.\vec{n} - \sin(2\pi - \theta).(\vec{x} \wedge \vec{n})$$

$$A_{\vec{n},2\pi-\theta}(\vec{x} \diamond \sigma) = \cos(-\theta).\vec{x} + (1 - \cos(-\theta)).\langle\vec{x},\vec{n}\rangle.\vec{n} - \sin(-\theta).(\vec{x} \wedge \vec{n}) = A_{\vec{n},-\theta}(\vec{x} \diamond \sigma)$$

proving that $R_{\vec{n},2\pi-\theta} = R_{\vec{n},-\theta} = R_{-\vec{n},\theta}$

Any $V_{\vec{n},\theta} \in SU(2)$ with $\theta \in [0; \pi[$ defines a rotation $R_{\vec{n},\theta} \in SO(3)$ (see Fig. 8) through the angle $\theta$ about the direction $\vec{n}$ which is the first cover of $SO(3)$, and $V_{\vec{n},\theta} \in SU(2)$ with $\theta \in [\pi; 2\pi[$ is the second cover of $SO(3)$.

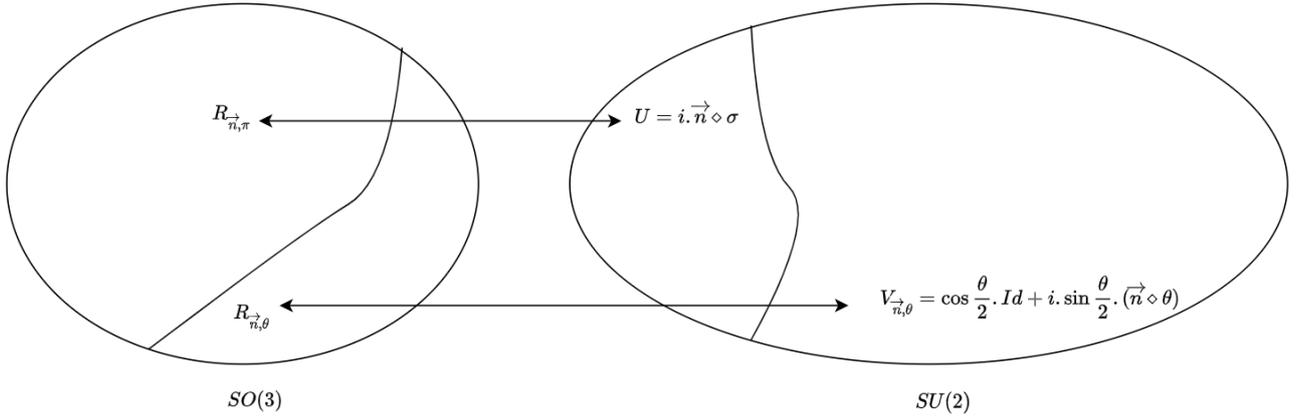

**Fig. 8.** The operators of $SU(2)$ mapped to rotations

Note that $i.(\vec{n} \diamond \sigma) = V_{\vec{n},\pi}$ is mapped to a rotation through the angle $\pi$ about the direction $\vec{n}$ that is a specific situation where $\cos\frac{\theta}{2} = 0$.

**Property.** The application $A: V_{\vec{n},\theta} \mapsto A(V_{\vec{n},\theta})$ is an epimorphism (surjective homomorphism) from $SU(2)$ onto $SO(3)$.

**Proof.**

For any $\vec{x} \in S^2$:

$$A(V_{\vec{n},\theta}.V_{\vec{m},\alpha})(\vec{x}) = V_{\vec{n},\theta}.\left(V_{\vec{m},\alpha}.i.(\vec{x} \diamond \sigma).\left(V_{\vec{m},\alpha}\right)^{\dagger}\right).\left(V_{\vec{n},\theta}\right)^{\dagger}$$

$$A(V_{\vec{n},\theta}.V_{\vec{m},\alpha})(\vec{x}) = V_{\vec{n},\theta}.i.\left(R_{\vec{m},\alpha}(\vec{x}) \diamond \sigma\right).\left(V_{\vec{n},\theta}\right)^{\dagger}$$

$$A(V_{\vec{n},\theta}.V_{\vec{m},\alpha})(\vec{x}) = i.\left(R_{\vec{n},\theta}.R_{\vec{m},\alpha}\right)(\vec{x}) \diamond \sigma$$

□

We have $V_{\vec{n},\theta}.V_{\vec{m},\alpha}$ that is mapped to $R_{\vec{n},\theta}.R_{\vec{m},\alpha}$ and we have a surjective homomorphism between $SU(2)$ and $SO(3)$. Because both $V_{\vec{n},\theta}$ and $V_{\vec{n},2\pi-\theta}$ are mapped to $R_{\vec{n},2\pi-\theta} = R_{\vec{n},-\theta}$, it is not an isomorphism between $SU(2)$ and $SO(3)$ but defines an isomorphism between $SO(3)$ and $SU(2)/\{Id; -Id\}$ that is identifying $V_{\vec{n},\theta}$ and $V_{\vec{n},2\pi-\theta}$.

## 2.5. Rotations

In rotation through the angle $\theta$ about the direction $\vec{n}$ is defined by:

$$V_{\vec{n},\theta} = R_{\vec{n},\theta} = e^{i.\frac{\theta}{2}\vec{n}\diamond\vec{\sigma}} = \cos\left(\frac{\theta}{2}\right).Id + i.\sin\left(\frac{\theta}{2}\right).\vec{n} \diamond \vec{\sigma}$$

A rotation through the angle $\theta$ about the direction $\vec{n} = (1; 0; 0)$ i.e. the $x$-axis is defined by:

$$R_{(1;0;0),\theta} = e^{-i.\frac{\theta}{2}.X} = \cos\left(\frac{\theta}{2}\right)Id + i\sin\left(\frac{\theta}{2}\right).X$$

referred to as (for convenience) $R_{X,\theta}$.

$$R_{X,\theta} = \begin{pmatrix} \cos\left(\frac{\theta}{2}\right) & i.\sin\left(\frac{\theta}{2}\right) \\ i.\sin\left(\frac{\theta}{2}\right) & \cos\left(\frac{\theta}{2}\right) \end{pmatrix}$$

Similarly, we have:

$$R_{Y,\theta} = e^{i.\frac{\theta}{2}.Y} = \cos\left(\frac{\theta}{2}\right).Id + i.\sin\left(\frac{\theta}{2}\right).Y = \begin{pmatrix} \cos\left(\frac{\theta}{2}\right) & \sin\left(\frac{\theta}{2}\right) \\ -\sin\left(\frac{\theta}{2}\right) & \cos\left(\frac{\theta}{2}\right) \end{pmatrix}$$

and

$$R_{Z,\theta} = e^{i.\frac{\theta}{2}.Z} = \cos\left(\frac{\theta}{2}\right).Id + i.\sin\left(\frac{\theta}{2}\right).Z = \begin{pmatrix} e^{i.\frac{\pi}{2}} & 0 \\ 0 & e^{-i.\frac{\pi}{2}} \end{pmatrix}$$

Remarks

$$R_{X,\pi} = e^{i.\frac{\pi}{2}.X} = i.X \in SU(2)$$

$$R_{Y,\pi} = e^{i.\frac{\pi}{2}.Y} = i.Y \in SU(2)$$

$$R_{Z,\pi} = e^{i.\frac{\pi}{2}.Z} = i.Z \in SU(2)$$

A rotation through the angle $\theta$ can be defined by two axial symmetries about the direction $\vec{n}$ and $\vec{m}$ with $\vec{n}.\vec{m} = \cos\theta/2$. $i.(\vec{n} \diamond \sigma) = V_{\vec{n},\pi}$ operators belong to a subset of $SU(2)$ since $i.(\vec{n} \diamond \sigma) = 0.Id + i.(\vec{n} \diamond \sigma)$ and these operators are composed of $i.X, i.Y$ and $i.Z$ i.e. of axial symmetries proving that any operator of $SU(2)$ can be defined by combinations of two axial symmetries.

**Definition**

For any two vectors $\vec{n}$ and $\vec{m}$ with $(\vec{n}, \vec{m})_{\vec{u}} = \frac{\theta}{2}$ about the direction $\vec{u} = \frac{\vec{n} \wedge \vec{m}}{\|\vec{n} \wedge \vec{m}\|}$ we have:

$$i.(\vec{m} \diamond \sigma).i.(\vec{n} \diamond \sigma) = V_{\vec{m},\pi}.V_{\vec{n},\pi} \text{ (Fig. 9)}$$

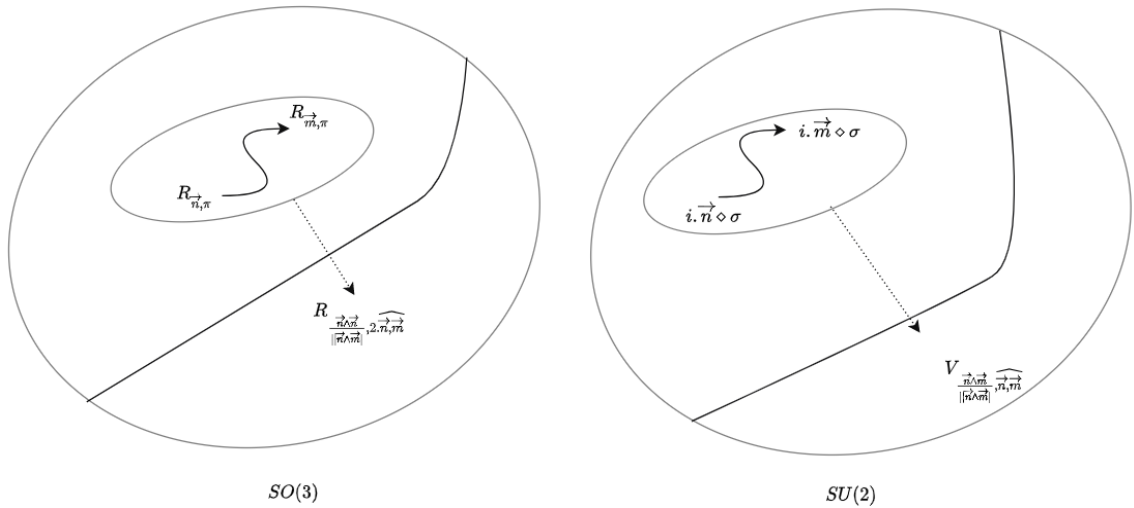

**Fig. 9.** Rotation through the angle $\theta$ about the direction $\vec{u}$

We have
$$A[i.(\vec{m} \diamond \sigma).i.(\vec{n} \diamond \sigma)] = A[V_{\vec{m},\pi}.V_{\vec{n},\pi}] = A[V_{\vec{m},\pi}].A[V_{\vec{n},\pi}] \sim R_{\vec{m},\pi}.R_{\vec{n},\pi} = R_{\vec{u},\theta} \sim A[V_{\vec{u},\theta}]$$
proving that
$$i.(\vec{m} \diamond \sigma).i.(\vec{n} \diamond \sigma) = V_{\vec{u},\theta} \text{ with } \theta \in [0; 2.\pi[$$

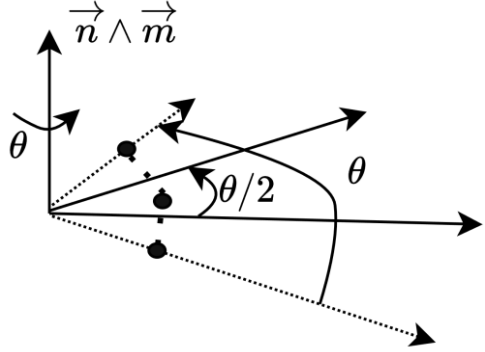

**Fig. 10.** Graphical representation of the rotation

$R_{\vec{m},\pi}.R_{\vec{n},\pi}$ is the successive application of two axial symmetries (Fig. 10) about the direction $\vec{n}$ first and $\vec{m}$ second with $\widehat{\vec{n},\vec{m}} = \frac{\theta}{2}$ that defines a rotation of angle $\theta$ about the direction $\vec{u} = \frac{\vec{n} \wedge \vec{m}}{\|\vec{n} \wedge \vec{m}\|}$.

For any vector $\vec{n} \in S^2$, we have $\vec{n} \diamond \sigma = (n_x.X + n_y.Y + n_z.Z)$ and $U(\vec{n},\sigma) = i.\vec{n} \diamond \vec{\sigma} \in SU(2)$ meaning that:
$$[U(\vec{n};\sigma)].[U(\vec{n};\sigma)]^\dagger = Id$$

With
$$i.\vec{n} \diamond \sigma = n_x.i.X + n_y.i.Y + n_z.i.Z$$
$$[i.U(\vec{n};\sigma)]^\dagger = -n_x.i.X - n_y.i.Y - n_z.i.Z$$

So
$$[i.U(\vec{n};\sigma)].[i.U(\vec{n};\sigma)]^\dagger = [n_x.i.X + n_y.i.Y + n_z.i.Z].[-n_x.i.X - n_y.i.Y - n_z.i.Z]$$
$$[i.U(\vec{n};\sigma)].[i.U(\vec{n};\sigma)]^\dagger = +n_x^2.Id + n_y^2.Id + n_z^2.Id$$
$$+n_x.n_Y.X.Y + n_x.n_z.X.Z + n_y.n_z.Y.Z$$
$$+n_x.n_Y.Y.X + n_x.n_z.Z.X + n_y.n_z.Z.Y$$

We have
$$X.Y = i.Z$$
$$Y.X = -i.Z$$
$$X.Z = -i.Y$$
$$Z.X = i.Y$$
$$Y.Z = i.X$$

$$Z.Y = -i.X$$

And by consequence $X.Y + Y.X = X.Z + Z.X = Y.Z + Z.Y = 0$.

So $[i.U(\vec{n};\sigma)].[i.U(\vec{n};\sigma)]^\dagger$ can be rewritten:

$$[i.U(\vec{n};\sigma)].[i.U(\vec{n};\sigma)]^\dagger = (n_x^2 + n_y^2 + n_z^2).Id = Id$$

Any operator $V_{\vec{n},\theta} \in SU(2)$ can be written: $V_{\vec{n},\theta} = \cos\frac{\theta}{2}.Id + i.\sin\frac{\theta}{2}.(n_x.X + n_y.Y + n_z.Z)$

And by consequence:

$$\cos^2\frac{\theta}{2} + \sin^2\frac{\theta}{2}.(n_x.X + n_y.Y + n_z.Z)^2 = 1$$

$$\cos^2\frac{\theta}{2} + \sin^2\frac{\theta}{2}.(n_x^2 + n_y^2 + n_z^2) = 1$$

$$1 - \sin^2\frac{\theta}{2} + \sin^2\frac{\theta}{2}.(n_x^2 + n_y^2 + n_z^2) = 1$$

$$\sin^2\frac{\theta}{2}.(n_x^2 + n_y^2 + n_z^2 - 1) = 0$$

This implies that either $\sin\frac{\theta}{2} = 0$ and $V_{\vec{n},\theta} = \pm Id$ or $n_x^2 + n_y^2 + n_z^2 = 1$ which is compliant with the definition.

## 3. Concluding remarks

In this paper we discussed about the mathematical foundations of $SU(2)$ operators and $SO(3)$ to provide a thorough survey of main results. Considering connections between operators and rotations, we aim to provide useful interpretations for researchers. We introduced first the qubit definition and a short description of qubit including the Bloch sphere and second a mathematical demonstration that there is a epimorphism (surjective homomorphism) from $SU(2)$ onto $SO(3)$.